\documentclass[12pt]{article}

\usepackage{amsmath}
\usepackage{amsthm}

\def\fnote#1#2{\begingroup\def\thefootnote{#1}\footnote{#2}
    \addtocounter{footnote}{-1}\endgroup}

\def\Email{natsuume@physics.upenn.edu, makoto.natsuume@kek.jp}
\def\asano{asano@post.kek.jp}

\newcommand{\be}{\begin{equation}}
\newcommand{\ee}{\end{equation}}
\newcommand{\bea}{\begin{eqnarray}}
\newcommand{\eea}{\end{eqnarray}}

\newcommand{\eq}[1]{(\ref{eq:#1})}
\newcommand{\sect}[1]{Sec.~\ref{sec:#1}}
\newcommand{\app}[1]{App.~\ref{sec:#1}}

\newcommand{\tabl}[1]{Table~\ref{table:#1}}

\newcommand{\bra}[1]{\mbox{$\langle #1 |$}}
\newcommand{\ket}[1]{\mbox{$| #1 \rangle$}}
\newcommand{\norm}[2]{\mbox{$\langle #1 | #2 \rangle$}}
\newcommand{\nor}{\stackrel{\scriptscriptstyle\circ}{\scriptscriptstyle\circ}}

\newcommand{\sig}[1]{\mbox{sign}(#1)}
\newcommand{\dims}[1]{\mbox{dim}(#1)}
\newcommand{\tr}{\mbox{tr}\,}

\newcommand{\defi}{\stackrel{\rm def}{=}}

\newcommand{\cond}[1]{\quad\mbox{#1}}

\newcommand{\Htotal}{{\cal H}_{\rm total}}
\newcommand{\Hc}{\hat{\cal H}_{\rm c}}
\newcommand{\He}{\hat{\cal H}_{\rm e}}
\newcommand{\obs}{\rm obs}

\newcommand{\hN}{\hat{N}^{\rm g}}
\newcommand{\hQ}{\hat{Q}}
\newcommand{\ta}{\tilde{a}}
\newcommand{\tb}{\tilde{b}}
\newcommand{\hc}{\hat{c}}
\newcommand{\hX}{h^{0}}
\newcommand{\hK}{h^{K}}
\newcommand{\am}{a^0}
\newcommand{\ag}{a^{\rm g}}
\newcommand{\cX}{\hc}
\newcommand{\cK}{\hc^{K}}
\newcommand{\Lint}{L_{0}^{\rm osc}}
\newcommand{\Lm}[1]{L^{0}_{#1}}
\newcommand{\Lg}[1]{L^{\rm g}_{#1}}
\newcommand{\LK}[1]{L^{K}_{#1}}
\newcommand{\Gm}[1]{G^{0}_{#1}}
\newcommand{\Gg}[1]{G^{\rm g}_{#1}}
\newcommand{\GK}[1]{G^{K}_{#1}}

\newcommand{\bet}[1]{\beta_{#1}}
\newcommand{\gam}[1]{\gamma_{#1}}
\newcommand{\FockA}{{\cal F}_{``\cX=1"}}
\newcommand{\FockB}{{\cal F}(c_{-m}, \gam{-r})}

\newcommand{\ap}{\alpha'}

\newtheorem{lemma}{Lemma}[section]
\newtheorem{theorem}{Theorem}[section]

\def\IR{\relax{\rm I\kern-.18em R}}

\begin{document}

\begin{flushright}
        KEK-TH-872, UPR-1031-T\\
        hep-th/0303051\\
\end{flushright}

\vspace{18pt}

\begin{center}
{\large \bf The no-ghost theorem} \\ \vspace{4pt}
{\large \bf in curved backgrounds with a timelike $u(1)$:} \\ \vspace{4pt}
{\large \bf NSR string}

\vspace{16pt}
Masako Asano $^{1}$ \fnote{\dag}{\asano} and Makoto Natsuume $^{1,2}$ 
\fnote{*}{\Email; On leave of absence from KEK.}

\vspace{16pt}

{\sl $^{1}$ Theory Division\\
Institute of Particle and Nuclear Studies\\
KEK, High Energy Accelerator Research Organization\\
Tsukuba, Ibaraki, 305-0801 Japan}
\vspace{8pt}

{\sl $^{2}$ Department of Physics and Astronomy\\
University of Pennsylvania\\
Philadelphia, PA 19104-6396, USA}

\vspace{12pt}
{\bf ABSTRACT}

\vspace{12pt}
\begin{minipage}{4.8in}
It is well-known that the standard no-ghost theorem is valid as long as the background has the light-cone directions. We prove the no-ghost theorem for the NSR string when only the timelike direction is flat. This is done by the BRST quantization, using the technique of Frenkel, Garland and Zuckerman and our previous results for the bosonic string. The theorem actually applies as long as the timelike direction is written as a $u(1)$ SCFT.
\end{minipage}

\end{center}

\begin{flushright}
       BRST quantization, vanishing theorem\\
\end{flushright}

\vfill
\pagebreak

\baselineskip=16pt

\section{Introduction}

In the last decade or so, strings on curved backgrounds have been discussed widely in various contexts. Some recent examples are $AdS$/CFT dualities, string on {\it pp}-wave backgrounds \cite{Berenstein:2002jq,Metsaev:2002re}, and time-dependent orbifolds \cite{Khoury:2001bz,Balasubramanian:2002ry,Liu:2002ft,Liu:2002kb}. However, many discussion is limited to backgrounds with light-cone directions ({\it e.g.}, string on {\it pp}-wave backgrounds and time-dependent orbifolds). Otherwise, the technology is often limited to supergravities ({\it e.g.}, for $AdS$/CFT dualities).

This is due to the lack of the string theory on general backgrounds, especially the no-ghost theorem. As is well-known, string theory generally contains negative norm states (ghosts) from timelike oscillators. However, they do not appear as physical states. This is well-established for string theory in flat spacetime \cite{Brower:1972wj}-\cite{Figueroa-O'Farrill:1989hu}. When the background spacetime is curved, things are not clear though. Standard proofs of the no-ghost theorem requires light-cone directions, {\it i.e.}, $d \geq 2$ if the background is written as $ \IR^{1,d-1} \times K $, where $K$ is a unitary CFT. This is true both in the old covariant quantization (OCQ) and in the BRST quantization (\tabl{standard_proofs})%
\footnote{There are several attempts of the $d=1$ proof in the old covariant quantization. For some discussion, see our first paper \cite{Asano:2000fp}. We will also discuss them thoroughly in our future paper \cite{Asano:2003qb}.}. 



However, the source of ghosts is the timelike oscillators and the Fadeev-Popov ghosts. Thus, one would expect that no-ghost theorem is valid even for $d=1$ as long as the timelike direction is intact. In fact, in our previous paper \cite{Asano:2000fp}, we show the no-ghost theorem for $d \geq 1$ bosonic string based on the BRST quantization. We heavily used the previous results by Frenkel, Garland and Zuckerman (FGZ) \cite{Frenkel:1986dg}. The purpose of this paper is to extend the proof for the NSR string. 

The proof by FGZ is different from the others. For example, the standard BRST quantization assumes $d \geq 2$ in order to prove the ``vanishing theorem," {\it i.e.}, the BRST cohomology is trivial except at the zero ghost number. However, FGZ's proof of the vanishing theorem essentially does not require $d \geq 2$. Moreover, the power of the technique is not limited to the $d=1$ case. This scheme is especially interesting because it does not even require that the timelike direction be flat; it admits an extension to more general curved backgrounds. As an example, we will discuss $AdS_3$ case in a separate paper \cite{Asano:2003qb}. 

Unfortunately, these points have not been appreciated well. This may be partly because the proof requires some mathematical backgrounds. It is one of our purposes here to explain FGZ's proof to a broader audience in a more accessible manner.

In the next section, we briefly review the BRST quantization of the NSR string. Further details and our conventions are summarized in \app{appA} and \ref{sec:appB}. The reader who is familiar to the BRST quantization can directly go to the outline of our proof in \sect{outline}. The presentation of the proof below is slightly different from our earlier paper \cite{Asano:2000fp}, but the proof itself is very similar.




\begin{table}
\begin{center}
\begin{tabular}{lll}
\hline
Quantization	& Approaches	& Limitations \\
\hline
&& \\
OCQ		& DDF \cite{Brower:1972wj}	& flat spacetime only \\
		& \'{a} la Goddard-Thorn \cite{Goddard:1972iy}	& $d \geq 2$ \\
BRST	& \'{a} la Kato-Ogawa \cite{Kato:1983im}		& $d \geq 2$ \\
		& Asano-Natsuume \cite{Asano:2000fp}			& $d \geq 1$ \\
&& \\
\hline
\end{tabular}
\caption{Standard schemes for the no-ghost theorem.}
\label{table:standard_proofs}
\end{center}
\end{table}

\section{Preliminary and Outline}\label{sec:brst}

\subsection{The Assumptions}\label{sec:assumptions}

We make the following assumptions:
\begin{enumerate}
\item[(i).] Our world-sheet theory consists of $d$ free bosons $ X^{\mu} $ and fermions $\psi^{\mu}\; (\mu = 0,\cdots ,d-1)$ with signature $ (1, d-1) $ and a unitary SCFT $ K $ of central charge $ \hc_{K} = 10-d $ ($ \hc = 2c/3 $). Although we focus on the $d=1$ case below, the extension to $1 \leq d \leq 10$ is straightforward.
\item[(ii).] 
We assume that $K$ is unitary and that all states in $K$ lie in highest weight representations. From the Kac determinant (\app{appC}), the weight of a highest weight state has $\hK > 0$ in the Neveu-Schwarz (NS) sector and $\hK > \hc_{K}/16$ in the Ramond (R) sector. An example of $K$ is a compact unitary SCFT, where its spectrum is discrete and bounded below. Another example is the transverse SCFT.
%
\item[(iii).] The momentum of states is $k^{\mu} \neq 0$.
\end{enumerate}
In \app{appA} and \ref{sec:appB}, we summarize our notations and conventions.

\subsection{BRST Quantization}

From our assumptions, the total $L_{m}$ of the theory is given by 
\be
L_{m} = \Lm{m} + \Lg{m} + \LK{m},
\ee
where $\Lm{m}$, $\Lg{m}$, and $\LK{m}$ represent the Virasoro generators in the $\hc=1$ timelike sector, the FP-ghost sector, and the unitary SCFT $K$ sector, respectively. In particular,
\bea
L_{0} &=& \ap k^2 + \Lint \\
&=& \ap k^2 + N + \LK{0} + \am + \ag,
\label{eq:Hamiltonian}
\eea
%
%
where $N$ is the total level number and $\am$ ($\ag$) represents the normal ordering constant for the $\hc=1$ (FP-ghost) sector. 

We will call the total Hilbert space $\Htotal$. For the NS-sector, the physical state conditions are
\be
Q \ket{\mbox{phys}} = 0
\ee
as well as 
\be
b_{0} \ket{\mbox{phys}} = L_{0} \ket{\mbox{phys}} = 0.
\ee
The $L_{0}$-condition follows from 
$0 = \{ Q, b_{0} \} \ket{\mbox{phys}} = L_{0} \ket{\mbox{phys}}$.
In addition, in the R-sector we impose
\be
\bet{0} \ket{\mbox{phys}} = G_{0} \ket{\mbox{phys}} = 0.
\ee
As in the $L_{0}$-condition, the $G_{0}$-condition follows from 
$0 = [Q, \bet{0}] \ket{\mbox{phys}} = G_{0} \ket{\mbox{phys}}$.

Thus, we define the following subspaces of $\Htotal$:

\noindent{\it (i) NS-sector}
\begin{subequations}
\begin{eqnarray}
{\cal H} &=& \{ \phi \in \Htotal: b_{0} \phi = 0 \}, \\
\hat{\cal H} &=& {\cal H}^{L_0} 
	= \{ \phi \in \Htotal: b_{0} \phi = L_0 \phi = 0 \}.
\end{eqnarray}
\label{eq:Hilbert_NS} 
\end{subequations}

\noindent{\it (ii) R-sector}
\begin{subequations}
\begin{eqnarray}
{\cal H} &=& \{ \phi \in \Htotal: 
	b_{0} \phi = \bet{0} \phi = 0 \}, \\
\hat{\cal H} &=& {\cal H}^{L_0, G_0} 
	= \{ \phi \in \Htotal: 
	b_{0} \phi = \bet{0} \phi = L_0 \phi = G_0 \phi = 0 \}.
\end{eqnarray}
\label{eq:Hilbert_R} 
\end{subequations}
Here, $*^{L_0}$ denotes the $L_0$-invariant subspace: 
$ F^{L_0} = F \cap \mbox{Ker} L_0$. (Similarly for $*^{L_0, G_0}$) We will consider the cohomology on $\hat{\cal H}$ since $Q$ takes $\hat{\cal H}$ into itself from $ \{Q, b_0\} = L_0 $ and $ [Q, L_0] = 0 $. (For the R-sector, also use $ [Q, \bet{0}] = G_0 $ and $ \{Q, G_0\} = 0 $.) The subspace ${\cal H}$ will be useful in our proof of the vanishing theorem (\sect{vanishing}).

The Hilbert space $\hat{\cal H}$ is classified according to mass eigenvalues. $\hat{\cal H}$ at a particular mass level will be often written as $\hat{\cal H}(k^2)$. For a state $ \ket{\phi} \in \hat{\cal H}(k^2) $,
\begin{equation}
L_{0} \ket{\phi} = (\alpha' k^2 + \Lint) \ket{\phi} = 0.
\label{eq:on_shell}
\end{equation}
One can further take an eigenstate of the ghost number $ \hN $ since $ [ \Lint, \hN ] = 0 $. $\hat{\cal H}$ is decomposed by the eigenvalues of $\hN$ as 
\begin{equation}
\hat{\cal H} = \bigoplus_{n\in {\bf Z}}\hat{\cal H}^n.
\end{equation}

We decompose the BRST operator $Q$ in terms of superconformal ghost zero modes:
\be
Q = \hQ
+ (\mbox{terms in $Q$ with superconformal ghost zero modes}).
        \label{eq:hatq}
\ee
See \app{appA} for the explicit form of $\hQ$. Then, for a state $\ket{\phi} \in \hat{\cal H}$,
\begin{equation}
Q \ket{\phi} = \hQ \ket{\phi}.
        \label{eq:relative}
\end{equation}
Therefore, the physical state condition reduces to
\begin{equation}
\hQ \ket{\phi} = 0.
\end{equation}
Also, $\hQ^2=0$ on $\hat{\cal H}$ from Eq.~\eq{relative}. Thus, $\hQ: \hat{\cal H}^n \rightarrow \hat{\cal H}^{n+1} $ defines a BRST complex, which is called the {\it relative} BRST complex. So, we can define $\Hc, \He \subset \hat{\cal H}$ by
\begin{equation}
\hQ \Hc = 0, \qquad
\He = \hQ \hat{\cal H},
\end{equation}
and define the relative BRST cohomology of $Q$ by
\begin{equation}
\hat{\cal H}_{\obs}=\Hc/\He.
\end{equation}
In terms of the cohomology group,
$\hat{\cal H}_{\obs}(k^2)=\oplus_{n \in {\bf Z}}\, H^n(\hat{\cal
H}(k^2), \hQ(k))$.

\subsection{The Outline of the Proof}\label{sec:outline}

The full proof of the no-ghost theorem is rather involved, so we give the outline here. The terminology appeared below is explained later. In general, the proof of the no-ghost theorem consists of 2 steps in the BRST quantization (\tabl{outline}).

\begin{itemize}

\item Step~1:
The first is to show the {\it vanishing theorem}. The vanishing theorem states that the $\hQ$-cohomology is trivial except at the zero ghost number. This is done by choosing an appropriate (bounded) {\it filtration} for your BRST operator $\hQ$. A filtration allows us to use a simplified BRST operator $Q_{0}$ and we can first study the cohomology of $ Q_{0} $. If the $ Q_{0} $-cohomology is trivial, so is the $\hQ$-cohomology; this is the reason why the filtration is so useful. However, the particular filtration used in standard proofs is also part of the reason why $d \geq 2$ in those proofs. 

\item Step~2:
The second is to compute and compare the {\it index} and the {\it signature} of the cohomology group explicitly. If the index is equal to the signature, the no-ghost theorem holds provided the vanishing theorem is valid.

\end{itemize}

\noindent
Step~1 and 2 themselves consist of several steps. For our approach, these are explained in \sect{step1outline} and \ref{sec:step2outline}, respectively.

However, in our approach, the matter Virasoro generators themselves play a very important role, and it is useful to have an additional step:

\begin{itemize}
\item Step~0:
Write the matter Hilbert space in terms of products of two Verma modules, one for the $\hc=1$ SCFT and the other for the unitary SCFT $K$.
\end{itemize}

\noindent The step is useful particularly at Step~1.2 and is convenient when one discusses more general spacetime backgrounds.

\begin{table}
\begin{center}
\begin{tabular}{lc}
Step~0:& Matter Hilbert space via Verma modules \\
&$\downarrow$ \\
Step~1:& The vanishing theorem using filtration \\
&(reason why $d \geq 2$ in standard proofs) \\
&$\downarrow$ \\
Step~2:& The no-ghost theorem
\end{tabular}
\caption{The outline of the proof.}
\label{table:outline}
\end{center}
\end{table}

\section{Step~0: Hilbert Space via Verma Modules}

First step towards the no-ghost theorem is to map the timelike $\hc=1$ matter Fock space to Verma modules. This is essential for proving the vanishing theorem; in the language of FGZ, this means that the $\hc=1$ CFT is an ``${\cal L}_-$-free module," which is a prime assumption of the vanishing theorem  (Theorem~1.12 of \cite{Frenkel:1986dg}).
Moreover, expressing the Hilbert space in this general form is useful when one discusses CFTs other than the $ \IR^{1,0} \times K $ case. 

Let ${\cal V}(\hc, h)$ be a Verma module with highest weight $h$ and central charge~$\hc$. Then, 
\begin{lemma}
There is an isomorphism between a Verma module and the $\hc=1$ matter Fock space:
\begin{equation}
\sum_{s} {\cal F}(\alpha^{0}_{-m}, \psi^{0}_{-r};s, k^0) \cong {\cal V}(1, \hX) 
\cond{if $(k^0)^2>0$.}
\label{eq:verma}
\end{equation}
(The index $s$ and its sum is relevant only to the R-sector. See \app{appB} and \ref{sec:appC} for notations.)%
\footnote
{At this stage, the Verma module implicitly defines an inner product on ${\cal F}$, which is the timelike part of ${\cal H}$. However, in the R-sector, the real inner product on $\hat{\cal H}$ used in the no-ghost theorem is slightly different from the one by the Verma module (See App. B). This does not matter to the present discussion since here we only need to show that the states in the Verma module are linearly independent. We use the inner product on $\hat{\cal H}$ given in App. B for the no-ghost theorem. Similar remark also applies to the expression \eq{hilbert}.\label{ft:inner_product}
}
\label{lemma:verma}
\end{lemma}
\noindent 
The isomorphism is plausible from the defining formula of $\Lm{m}$ and $\Gm{r}$,
\bea
\Lm{-m} &=& \sqrt{2\alpha'} k_0 \alpha^{0}_{-m} + \cdots, 
\quad (m \neq 0) \\
\Gm{-r} &=& \sqrt{2\alpha'} k_0 \psi^{0}_{-r} + \cdots,
\eea
where $+ \cdots$ denotes terms with more than one oscillators. The above relations also suggest that the isomorphism fails at $k^0=0$. This is the reason why we require assumption~(iii) in \sect{assumptions}.

\begin{proof}[Proof of Lemma~\ref{lemma:verma}]
The number of states of the Fock space $\sum_s {\cal F}(\alpha^{0}_{-m}, \psi^{0}_{-r};s,k^0)$ and that of the Verma module ${\cal V}(1, \hX)$ are the same for a given level $N$. Thus, the Verma module 
furnishes a basis of the Fock space if all the states in a highest weight representation are linearly independent. This can be shown using the Kac determinant (\app{appC}). 
%
%
%
%
For $\hc=1$, the Kac determinant does not vanish if 
\be
\hX<0 (\mbox{NS}), 1/16 (\mbox{R}),
\label{eq:roots}
\ee
so the states in the Verma module are linearly independent. Now, note that $\hX = -\ap (k^0)^2 + \am$, where $\am = 0 (\mbox{NS}), 1/16 (\mbox{R})$. Thus, Eq.~\eq{roots} is valid for $(k^0)^2>0$.
\end{proof}

The isomorphism is valid as long as $(k^0)^2>0$. Let us check what on-shell states actually appear in $\hat{\cal H}$. From assumption~(ii) of \sect{assumptions} and the on-shell condition, 
\begin{subequations}
\be
-\hX (= \ap (k^0)^2 - \am) = \ag + \LK{0} + N,
\ee
where
\be
\mbox{NS}
	\left\{ 
	\begin{array}{l}
	\am = 0 \\
	\ag = -\frac{1}{2} \\
	\hK > 0
	\end{array}\right.,
\qquad
	\mbox{R}
	\left\{ 
	\begin{array}{l}
	\am = \frac{1}{16} \\
	\ag = -\frac{10}{16} \\
	\hK > \frac{9}{16}
	\end{array}\right..
\ee
\label{eq:onshell}
\end{subequations}
\noindent
In the NS-sector, $\ap (k^0)^2 > -1/2$ from Eqs.~\eq{onshell}. Also, $k^0 \neq 0$ from assumption~(iii). The Fock spaces with $(k^0)^2>0$ are expressed by Verma modules. Those with $0 > \ap (k^0)^2 >-1/2$ are not. However, there is no state in this region. In the R-sector, one always has $(k^0)^2>0$. 

To summarize, the matter part of the Hilbert space ${\cal H}$ reduces to a sum of two Verma modules:
\begin{equation}
%
{\cal H}_{\hX,\hK} = 
{\cal V}(\cX=1, \hX<0) \otimes {\cal V}(\cK=9, \hK>0)
\label{eq:hilbert}
\end{equation}
for the NS-sector, and similarly for the R-sector.%
\footnote{See footnote~\ref{ft:inner_product}.} 
Throughout the discussion of the vanishing theorem in \sect{vanishing}, we assume the above form of the Hilbert space. {\it Consequently, the vanishing theorem is valid not only for the $d=1$ case but also for more general backgrounds as long as the matter Hilbert space takes the above form.}

\section{Step~1: The Vanishing Theorem}\label{sec:vanishing}

\subsection{Filtration and the Vanishing Theorem: General Discussion}

We now state our vanishing theorem:

\begin{theorem}[The Vanishing Theorem for String Theory]
The $\hQ$-cohomology can be non-zero only at $ \hN =0 $, {\it i.e.}, 
\be
H^n(\hat{\cal H}, \hQ)=0 \quad \mbox{for} \quad n\neq 0
\ee
if the matter part of $ {\cal H} $ can be decomposed as a sum of two Verma modules as in Eq.~\eq{hilbert}. \label{thm:vanishing}
\end{theorem}
\noindent To prove this, the notion of {\it filtration} is useful. However, a particular filtration used in standard proofs is part of the reason why $d \geq 2$.

A filtration is a procedure to break up $\hQ$ according to a quantum number $ N_{f} $ (filtration degree):
\begin{equation}
\hQ = Q_{0}+Q_{1}+\cdots+Q_{N},
\end{equation}
where
\begin{equation}
[N_{f}, Q_{m}] = m Q_{m}.
\end{equation}
In order that $N_{f}$ takes $\hat{\cal H}$ into itself, a filtration also satisfies
\begin{equation}
[N_{f}, \hN] = [N_{f}, L_{0}] = [N_{f}, G_{0}] = 0.
\label{eq:cond}
\end{equation}
If $\hat{\cal H}$ can be nonzero only for a finite range of degrees, the filtration is called {\it bounded}.

The nilpotency of $ \hQ^2 $ implies
\begin{equation}
\sum_{\stackrel{\scriptstyle m, n}{m+n=l}}
Q_{m}Q_{n} = 0, \qquad l=0, \ldots ,2N
\end{equation}
since they have different values of $ N_{f} $. In particular,
\begin{equation}
Q_{0}^2 = 0.
\end{equation}
The point is that we can first study the cohomology of $ Q_{0} $. This is easier since $ Q_{0} $ is often simpler than $ \hQ$. Knowing the cohomology of $ Q_{0} $ then tells us about the cohomology of $ \hQ $. In fact, one can show the following lemma (for a bounded filtration):
\begin{lemma}
If the $ Q_{0} $-cohomology is trivial, so is the $ \hQ
$-cohomology.
\label{lemma:exact_sequence}
\end{lemma}
\noindent See, {\it e.g.}, Ref.~\cite{Asano:2000fp} for the proof. 

Note that the above lemma states only for trivial cohomology; nontrivial cohomology for $Q$ is in general different from the $Q_{0}$-cohomology. However, one can show that the $Q_{0}$-cohomology is isomorphic to that of $\hQ$ if the $Q_{0}$-cohomology is nontrivial for at most one filtration degree \cite{Polchinski:rq,Bouwknegt:1992yg}. Then, a standard proof proceeds to show that states in the nontrivial degree are in fact light-cone spectra, and thus there is no ghost in the $\hQ$-cohomology \cite{Polchinski:rq}. We will not take this path to prove the no-ghost theorem. However, the vanishing theorem is useful in our approach as well.

Now, we have to find an appropriate filtration and show that the $Q_0$-cohomology is trivial if $\hN \neq 0$. This completes the proof of the vanishing theorem. The standard proof of the theorem uses the following filtration \cite{Kato:1983im,Bouwknegt:1992yg,Polchinski:rq}: \footnote{The $\hN$ piece is not really necessary. We include this to make the filtration degree non-negative.}
\begin{equation}
N^{\rm (KO)}_{f} = \sum_{\stackrel{\scriptstyle m=-\infty}{m \neq 0}}^{\infty}
        \frac{1}{m} \alpha^{-}_{-m} \alpha^{+}_{m} + \hN.
\label{eq:Nko}
\end{equation}
The degree $N^{\rm (KO)}_{f}$ counts the number of $\alpha^+$ minus the number of $\alpha^-$ excitations. So, this filtration assumes two flat directions, and we have to take a different approach for $d=1$.

%
%


\subsection{The Outline of Step~1}\label{sec:step1outline}


Since we want to show the no-ghost theorem for $d=1$, we cannot use $ N^{\rm (KO)}_{f} $ as our filtration degree. Fortunately, the structure of our Hilbert space \eq{hilbert} enables us to prove the vanishing theorem using a different filtration \cite{Frenkel:1986dg,Lian:1989cy,Figueroa-O'Farrill:1989hu}. 
The proof of the vanishing theorem consists of three steps (\tabl{vanishing}):

\begin{itemize}
\item Step~1.1:
Apply our filtration \'{a} la Frenkel, Garland, and Zuckerman. With FGZ's filtration, $Q_0$ can be further decomposed as a sum of two differentials, $d'$ and $d''$. This decomposition is crucial for the proof; it effectively reduces the problem to a ``$\cX=1$" SCFT, which contains the timelike part, the $b$ and $\beta$ ghost part. This is the reason why the proof does not require $d \geq 2$. 
\item Step~1.2:
If the $d'$-cohomology is trivial, so is the $Q_0$-cohomology. This follows from a K\"{u}nneth formula. Then, the $\hQ$-cohomology is trivial as well from Lemma~\ref{lemma:exact_sequence}.
\item Step~1.3:
Now, the problem is reduced to the $d'$-cohomology. Show the vanishing theorem for the $d'$-cohomology. 
%
\end{itemize}

\noindent In this approach, the matter Virasoro generators themselves play a role similar to that of the light-cone oscillators in Kato-Ogawa's approach. In this section, we prove the theorem using the technique of Refs.~\cite{Frenkel:1986dg,Lian:1989cy,Figueroa-O'Farrill:1989hu}, but for more mathematically rigorous discussion, consult the original references.

\begin{table}
\begin{center}
\begin{tabular}{c}
Filtration (Step~1.1) \\
$\downarrow$ \\
$Q^{\rm (FGZ)}_{0} = d'+d''$ \\
effectively reduces the problem to a ``$\cX=1$" part \\
$\downarrow$ \\
The Vanishing Theorem for the $d'$-cohomology (Step~1.3) \\
$\downarrow$ \\
The Vanishing Theorem for the $Q^{\rm (FGZ)}_{0}$-cohomology (Step~1.2) \\
$\downarrow$ \\
The Vanishing Theorem for the $\hQ$-cohomology (Theorem~\ref{lemma:exact_sequence}) \\
\end{tabular}
\caption{The outline of the proof of the vanishing theorem for $d=1$.}
\label{table:vanishing}
\end{center}
\end{table}

%
%

\subsection{Step~1.1: Filtration}

Our filtration is given by 
\begin{equation}
N^{\rm (FGZ)}_{f} = -\Lm{0} + \sum_{m>0} m ( N_{m}^{c} -
N_{m}^{b} ) + \sum_{r>0} r ( N_{r}^{\gamma} -
N_{r}^{\beta} ).
\label{eq:fgz}
\end{equation}
The filtration assigns the degrees in \tabl{filtration1} to the operators. FGZ's filtration is originally given for the $d=26$ bosonic string, and it was later extended to the flat $d=10$ NSR string \cite{Lian:1989cy,Figueroa-O'Farrill:1989hu}. We can apply this filtration to our problem since it does not require $d \geq 2$ in principle.

\begin{table}
\begin{center}
\[
\begin{array}{cr}
\hline
\mbox{Operators} & \mbox{fdeg} \\
\hline
c_{m} 	& |m|	\\
b_{m} 	& -|m| 	\\
\gam{r}	& |r|	\\
\bet{r}	& -|r|	\\
\Lm{m}	& m		\\ 
\Gm{r}	& r		\\
\LK{m}	& 0		\\
\GK{r}	& 0		\\
\hline
\end{array}
\]
\caption{Filtration degrees to each modes for the vanishing theorem.}
\label{table:filtration1}
\end{center}
\end{table}

The operator $N^{\rm (FGZ)}_{f}$ satisfies conditions~\eq{cond} and the degree of each term in $\hQ$ is non-negative. Because the eigenvalue of $\Lint$ is bounded below from Eqs.~\eq{Hamiltonian} and \eq{on_shell}, the total number of oscillators for a given mass level is bounded. Thus, the degree for the states is bounded for each mass level. Note that the unitarity of $K$ is essential for the filtration to be bounded.

%
%
The degree zero part of $\hQ$ is given by
\begin{subequations}
\label{eq:q0}
\begin{eqnarray}
Q^{\rm (FGZ)}_{0} &=& d'+d'', \\
d' &=& \sum_{m>0} c_{m} \Lm{-m} + \sum_{r>0} \gam{r} \Gm{-r} 
        + \sum_{m,n>0} \frac{1}{2} (m-n) b_{-m-n} c_{m} c_{n}
\nonumber \\
     && + \sum_{m,r>0} \frac{1}{2} (2r-m) \bet{-m-r} c_{m} \gam{r} 
        - \sum_{m,r>0} b_{-m-r} \gam{m} \gam{r}, 
\\
d'' &=& - \sum_{m,n>0} \frac{1}{2} (m-n) c_{-m} c_{-n} b_{m+n} 
\nonumber \\
     && - \sum_{m,r>0} \frac{1}{2} (2r-m) \bet{m+r} c_{-m} \gam{-r}
        - \sum_{m,r>0} b_{m+r} \gam{-m} \gam{-r}.
\end{eqnarray}
\end{subequations}
Note that $d'$ ($d''$) includes only $c_{m>0}$ and $\gam{r>0}$ ($b_{m>0}$ and $\bet{r>0}$). Also, the matter part is included in $d'$ only. 
The importance of the filtration is that $Q^{\rm (FGZ)}_{0}$ naturally breaks up into two differentials $d'$ and $d''$ further. To see this, break ${\cal H}$ as follows:
\begin{subequations}
\begin{eqnarray}
{\cal H} 
& = & 
{\cal V}(1, \hX) \otimes 
{\cal F}(b_{-m}, c_{-m}, \bet{-r}, \gam{-r}) \otimes 
{\cal H}_{K} \\
& = & 
\FockA \otimes \FockB \otimes {\cal H}_{K}.
%
\end{eqnarray}
\end{subequations}
Here, 
$\FockA = {\cal V}(1, \hX) \otimes {\cal F}(b_{-m}, \bet{-r}) $.
The Hilbert spaces ${\cal H}$, $\FockA$, and $\FockB$ are decomposed according to the ghost number
$\hN=n$:
\begin{equation}
{\cal H}^{n} =
\Bigl(
\bigoplus_{\stackrel{\scriptstyle n=c-b}{c, b \geq 0}}
\FockA^{-b} \otimes {\cal F}^{c}(c_{-m}, \gam{-r})
\Bigr) \otimes {\cal H}_{K},
%
\end{equation}
where
\begin{subequations}
\bea
b &=& N^{b} + N^{\beta}, \\
c &=& N^{c} + N^{\gamma}.
\eea
\end{subequations}
{}Then, the differentials act as follows:
\begin{subequations}
\begin{eqnarray}
&& Q^{\rm (FGZ)}_{0}: {\cal H}^n \rightarrow {\cal H}^{n+1}, \\
&& d': \FockA^{n} \rightarrow \FockA^{n+1}, 
\\
&& d'': {\cal F}^{n}(c_{-m}, \gam{-r}) 
\rightarrow 
{\cal F}^{n+1}(c_{-m}, \gam{-r}),
\end{eqnarray}
\end{subequations}
and $d'{}^2 = d''{}^2 = 0$. Thus, 
$\FockA^{n}$ and ${\cal F}^{n}(c_{-m}, \gam{-r})$ 
are complexes with differentials $d'$ and $d''$. Note that $Q^{\rm (FGZ)}_{0}$ is the differential for ${\cal H}^n$ as well as for $\hat{\cal H}^{n}$. We consider the cohomology on ${\cal H}^n$ for the time being, but eventually relate it to the cohomology on $\hat{\cal H}^{n}$.

\subsection{Step~1.2: Reduce the Problem to the $d'$-cohomology}

The property that $Q^{\rm (FGZ)}_{0}$ is the sum of two differentials $d'$ and $d''$ has an important consequence. This reduces the problem to the ``$\cX=1$" part only; we show that the vanishing theorem holds for the $Q^{\rm (FGZ)}_{0}$-cohomology 
{\it if} the theorem holds for the $d'$-cohomology. 
Then, in the next subsection, we see that this assumption for $d'$-cohomology certainly holds.

First, 
decompose a $Q^{\rm (FGZ)}_{0}$-closed state $\phi^n$ into 
a sum of products of $\phi_1^{-b}$ and $\phi_2^{n+b}$
with $b \geq 0$ and $n+b \geq 0$:
\be
\phi^n = \sum_{b, n+b \geq 0} \phi_1^{-b} \phi_2^{n+b}. 
\label{eq:kunneth1}
\ee
The superscripts denote their ghost numbers. Suppose that every $\phi_1$ is $d'$-exact for nonzero $b$:
\be
\phi_1^{-b} = d' \chi_1^{-b-1}, \qquad b>0.
\ee
%
%
As a consequence, for $\phi^n$ with $n<0$, 
\bea
\phi^n &=& \sum_{b \geq -n>0} \phi_1^{-b} \phi_2^{n+b}
\nonumber \\
&=& \sum  d' \chi_1^{-b-1} \phi_2^{n+b}.
\eea
Then, $\phi_2$ is $d''$-closed since
\be
0 
= Q^{\rm (FGZ)}_{0} \phi^n 
= \sum d' \chi_1^{-b-1} (-)^b (d'' \phi_2^{n+b}). 
\ee
So, each term in the sum \eq{kunneth1} can be written as
\bea
\phi _1 \phi_2
&=& (d' \chi_1) \phi_2
\nonumber \\
&=& (d' \chi_1) \phi_2 + (-)^{b+1} \chi_1 d'' \phi_2
\nonumber \\ 
&=& Q^{\rm (FGZ)}_{0} (\chi_1 \phi_2) \cond{for $n<0$}.
\label{eq:kunneth2}
\eea
Thus, $\phi$ is in fact $Q^{\rm (FGZ)}_{0}$-exact. To summarize, we have shown that
\begin{center}
If $\phi_1^{-b}$ is $d'$-exact for $b>0$, $\phi^n$ is $Q^{\rm (FGZ)}_{0}$-exact for $n<0$.
\end{center}

We can understand 
this as a consequence of a K\"{u}nneth formula. The K\"{u}nneth formula relates the cohomology group of ${\cal H}$ to those of $\FockA$ and $\FockB$:
\be
H^{n}({\cal H}) =
\bigoplus_{\stackrel{\scriptstyle n=c-b}{c, b \geq 0}}
	H^{-b}\left(\FockA\right)
	\otimes
	H^{c}\left(\FockB\right)
\otimes {\cal H}_K.
%
\ee
If $ H^{-b}\left(\FockA\right) = 0 $ for $b > 0$,
\be
H^{n}({\cal H}) =
	\bigoplus_{n=c }
	H^{0}\left(\FockA\right)
	\otimes
	H^{c}\left({\cal F}(c_{-m}^{}, \gam{-r}^{})\right)
	\otimes {\cal H}_K,
\label{eq:kunneth}
\ee
which leads to $H^{n}({\cal H}) = 0$ for $n<0$ because $c\geq 0$. Then, $H^{n}({\cal H})^{L_0} = 0$ for $n<0$.

The cohomology group we need is $H^{n}(\hat{\cal H})$, not $H^{n}({\cal H})^{L_0}$. However, Lian and Zuckerman have shown that 
\begin{equation}
H^n({\cal H}^{L_0}) \cong H^n({\cal H})^{L_0}.
\label{eq:LZ}
\end{equation}
For the R-sector, $\hat{\cal H}={\cal H}^{L_0, G_0}$. From the above result, it can be also shown that $H^n({\cal H}^{G_0})=0$ for $n<0$. See pages 325-326 of Ref.~\cite{Lian:1989cy}. Thus,
\begin{equation}
H^{n}(\hat{\cal H}, Q^{\rm (FGZ)}_{0}) = 0 
\cond{if $n<0$,}
\end{equation}
and using Lemma~\ref{lemma:exact_sequence}, 
\begin{equation}
H^{n}(\hat{\cal H}, \hQ) = H^{n}(\hat{\cal H}, Q^{\rm (FGZ)}_{0}) = 0 
\cond{if $n<0$.}
\end{equation}
We will later prove the Poincar\'{e} duality theorem,
$ H^{n}(\hat{\cal H}, \hQ) = H^{-n}(\hat{\cal H}, \hQ) $
(Lemma~\ref{lemma:poincare}). Therefore,
\begin{equation}
H^{n}(\hat{\cal H}, \hQ) = 0 \cond{if $n \neq 0$.}
\end{equation}
This is our vanishing theorem.

Actually, from Eq.~\eq{kunneth}, $H^{n}({\cal H})$ does not include a state with $b, c \ne 0$. Thus, we have established a stronger statement:
\begin{theorem}[FP-Ghost Decoupling Theorem]
Physical states do not contain Fadeev-Popov ghosts if $ {\cal H} $ is decomposed as in Eq.~\eq{hilbert}.
\end{theorem}
\noindent
Although the theorem itself is not necessary to establish the no-ghost theorem, it is useful to establish, {\it e.g.}, the BRST-OCQ equivalence \cite{Asano:2003qb}. 

To summarize, the problem of the $\hQ$-cohomology is reduced to the one of the $Q^{\rm (FGZ)}_{0}$-cohomology by Lemma~\ref{lemma:exact_sequence}, and we see in this subsection that the problem is further reduced to the one of the $d'$-cohomology. Thus, our problem now is
\begin{lemma}[The Vanishing Theorem for the ``$\cX=1$" Part]
The $d'$-cohomology can be non-zero only at zero ghost number, {\it i.e.}, $ H^{-b}\left(\FockA\right) = 0 $ if
$b > 0$.
\label{lemma:vanishingforc=1}
\end{lemma}
\noindent We prove the lemma in the next subsection.

\subsection{Step~1.3: The Vanishing Theorem for the $d'$-cohomology}\label{sec:vanishingforc=1}

We now show the vanishing theorem for the ``$\cX=1$" part (Lemma~\ref{lemma:vanishingforc=1}). The proof is straightforward using Step~0 and an argument given in \cite{Figueroa-O'Farrill:1989hu}.

\begin{proof}[Proof of Lemma~\ref{lemma:vanishingforc=1}]
{}From Step~0, a state $\ket{\phi} \in \FockA$ can be written as
\begin{equation}
\ket{\phi} = \bet{-r_{1}} \ldots \bet{-r_{K}} b_{-i_{1}} \ldots b_{-i_{L}} 
	\Gm{-\gam{1}} \ldots \Gm{-\gam{N}} 
	\Lm{-\lambda_{1}} \ldots \Lm{-\lambda_{M}}
	\ket{\hX},
\label{eq:statebn}
\end{equation}
where
\bea
&& 0 < r_{1} \leq r_{2} \leq \cdots \leq r_{K}, \nonumber \\
&& 0 < i_{1} < i_{2} < \cdots < i_{L}, \nonumber \\
&& 0 \leq \gam{1} < \gam{2} < \cdots < \gam{N}, \nonumber \\
&&0 < \lambda_{1} \leq \lambda_{2} \leq \cdots \leq \lambda_{M}.
\eea
Note that the states in $\FockA$ all have nonpositive ghost number: $\hN \ket{\phi} = -(K+L) \ket{\phi}$.

\begin{table}
\begin{center}
\[
\begin{array}{cr}
\hline
\mbox{Operators} & \mbox{fdeg} \\
\hline
c_{m} 		& 1		\\
b_{-m} 		& -1 	\\
\gam{r}		& 1		\\
\bet{-r}	& -1	\\
\Lm{-m}		& -1	\\ 
\Gm{-r}		& -1	\\	
\hline
\end{array}
\]
\caption{Filtration degrees to each modes (for $m, r>0$) for Lemma~\protect\ref{lemma:vanishingforc=1}.}
\label{table:filtration2}
\end{center}
\end{table}

We define a new filtration degree $N^{\rm (FK)}_{f}$ as
\be
N^{\rm (FK)}_{f} \ket{\phi} = -(K+L+M+N) \ket{\phi},
\ee
which corresponds to the assignments in \tabl{filtration2}. 
The operator $N^{\rm (FK)}_{f}$ satisfies conditions~\eq{cond}. 
Note that the filtration is not ``consistent" with the Virasoro algebra: if one applies the rule to $[\Lm{m}, \Lm{n}] = (m-n) \Lm{m+n} + \cdots $, the degree of the left-hand side is $-2$, whereas the degree of the right-hand side is $-1$. This means that the filtration degree of a state $\ket{\phi}$ can be determined only {\it after} one specifies the ordering of the Virasoro generators. 
Here, we take the above ordering. Similarly, $d'$ cannot be decomposed in itself; 
the degrees of $d'$ are determined after one chooses a state $\ket{\phi}$ and arranges 
$d' \ket{\phi}$ in the above ordering. 
In general, the degree of $d'$ is always non-negative, so we would like to extract the lowest degree, the degree zero part $d'_{0}$ of $d'$. Note that
\be
d' \ket{\phi} = 
\left( \sum_{m>0} c_{m} \Lm{-m} + \sum_{r>0} \gam{r} \Gm{-r} \right) 
\ket{\phi}+(N^{\rm (FK)}_{f}>0 \mbox{ terms}).
\ee
So, the $d'_{0}$-part comes only from the first term. The first term may still include $N^{\rm (FK)}_{f}>0$ terms; the $d'_{0}$-part can be extracted only after one arranges the first term in the above ordering. In practice, this is easy and one just has to put operators in the correct position. Since Virasoro generators do not commute, extra terms may appear, but these extra terms do not contribute to the degree zero part. %
\footnote{This paragraph corrects a misleading argument in the analogous proof for the bosonic string (Lemma~4.1) in Ref.~\cite{Asano:2000fp}.}
%
%

Since we want a bounded filtration, break up $\FockA$ according to $L_0$ eigenvalue $l_0$ :
\begin{equation}
\FockA = \bigoplus_{l_0} \FockA^{l_0},
\end{equation}
where
\begin{equation}
\FockA^{l_0}
 = \FockA \cap\mbox{Ker}(L_0-l_0).
\end{equation}
Then, the above filtration is bounded for each $\FockA^{l_0}$ since $\FockA^{l_0}$ is finite dimensional.

%
%

We first consider the $d'_{0}$-cohomology on $\FockA^{l_0}$ for each $l_0$. Define an operator $\Gamma$ such as
\bea
\lefteqn{\Gamma \ket{\phi} =} 
\nonumber \\
&& \sum_{l=1}^{M} \bet{-r_{1}} \ldots \bet{-r_{K}} 
b_{-\lambda_{l}} \left( b_{-i_{1}} \ldots b_{-i_{L}} \right)
\Lm{-\lambda_{1}} \ldots \widehat{\Lm{-\lambda_{l}}} \ldots
\Lm{-\lambda_{M}} 
\Gm{-r_1} \ldots \Gm{-r_N} \ket{\hX} 
\\
&& + \sum_{l=1}^{N} 
(-)^{L+l-1}
\bet{-r_{l}} \left( \bet{-r_{1}} \ldots \bet{-r_{K}} \right)
b_{-\lambda_{l}} \ldots b_{-i_{L}}
\Lm{-\lambda_{1}} \ldots \Lm{-\lambda_{M}} 
\Gm{-r_1} \ldots \widehat{\Gm{-r_{l}}} \ldots \Gm{-r_N} \ket{\hX},
\nonumber
\eea
where $\widehat{\Lm{-\lambda_{l}}}$ and $\widehat{\Gm{-r_{l}}}$ mean that the term is missing (When $M=0$ or $N=0$, $\Gamma \ket{\phi} \defi 0$). Then, it is straightforward to show that
\begin{equation}
\{ d'_{0}, \Gamma \} \ket{\phi} = (K+L+M+N) \ket{\phi}.
\end{equation}
The operator $\Gamma$ is called a {\it homotopy operator} for $d'_{0}$. Its significance is that the $d'_{0}$-cohomology is trivial except for $K+\cdots+N = 0$. If $\ket{\phi}$ is closed, then
\begin{equation}
\ket{\phi}      = \frac{\{ d'_{0}, \Gamma \}}{K+\cdots+N} \ket{\phi}
                = \frac{1}{K+\cdots+N} d'_{0} \Gamma \ket{\phi}.
\end{equation}
Thus, a closed state $\ket{\phi}$ is actually an exact state if $K+\cdots+N\neq 0$. Therefore, the $d'_{0}$-cohomology is trivial for $\hN<0$ since $\hN=-(K+L)$. And now, again using Lemma~\ref{lemma:exact_sequence}, the $d'$-cohomology $H^n(\FockA^{l_0})$ is trivial if $n<0$.

Because $[d',L_0]=0$, we can define
\begin{equation}
H^n(\FockA)^{l_0} =
H^n(\FockA)\cap \mbox{Ker}(L_0-l_0).
\end{equation}
Furthermore, as in Eq.~\eq{LZ}, the isomorphism
\begin{equation}
H^n(\FockA)^{l_0} \cong
H^n(\FockA^{l_0})
\end{equation}
can be established. Consequently, 
$H^n(\FockA)=0$ if $n <0$.
\end{proof}

\section{Step~2: The No-Ghost Theorem}\label{sec:no-ghost}

Having shown the vanishing theorem, it is straightforward to show the no-ghost theorem:
\begin{theorem}[The No-Ghost Theorem]
$\hat{\cal H}_{\obs}$ is a positive definite space when $1 \leq d \leq 10$. 
%
\end{theorem}
\noindent The calculation below is essentially the same as the one in Refs.~\cite{Lian:1989cy,Figueroa-O'Farrill:1989hu}, but we repeat it here for completeness.

\subsection{The Outline of Step~2}\label{sec:step2outline}

In order to prove the theorem, the notion of {\it signature} is useful. For a vector space $V$ with an inner product, we can choose a basis $e_{a}$ such that
\begin{equation}
\norm{e_{a}}{e_{b}} = \delta_{ab} C_{a},
\label{eq:basis}
\end{equation}
where $C_{a}\in \{0, \pm 1 \}$. Then, the signature of $V$ is defined as
\begin{equation}
\sig{V} = \sum_{a} C_{a},
\end{equation}
which is independent of the choice of $e_{a}$. Note that if $ \sig{V} = \dims{V} $, all the $C_{a}$ are 1, so $V$ has positive definite norm.

So, the statement of the no-ghost theorem is equivalent to 
\footnote{In this section, we also write
$V^{\obs}_i = \hat{\cal H}_{\obs}(k^2)$ and $V_i =\hat{\cal H}(k^2)$, where the subscript $i$ labels different mass levels.}
\begin{equation}
\sig{V^{\obs}_{i}} = \dims{V^{\obs}_{i}}.
        \label{eq:noghost1}
\end{equation}
This can be replaced as a more useful form
\begin{equation}
\sum_{i} e^{-\lambda \alpha' m_i^2} \sig{V^{\obs}_{i}}
        = \sum_{i} e^{-\lambda \alpha' m_i^2} \dims{V^{\obs}_{i}},
        \label{eq:noghost2}
\end{equation}
where $\lambda$ is a constant or
\begin{equation}
\mbox{tr}_{\obs} \, q^{\Lint} C
        = \mbox{tr}_{\obs} \, q^{\Lint},
\label{eq:noghost3}
\end{equation}
where $q=e^{-\lambda}$ and we have used the on-shell condition \eq{on_shell}. The operator $C$ gives eigenvalues $C_{a}$.

Equation~\eq{noghost3} is not easy to calculate; however, the following relation is straightforward to prove:
\begin{equation}
\mbox{tr} \, q^{\Lint} C
        = \mbox{tr} \, q^{\Lint} (-)^{\hN}.
\label{eq:step3}
\end{equation}
Here, the trace is taken over $V_{i}$ and we take a basis which diagonalizes $(-)^{\hN}$. Then, we can prove Eq.~\eq{noghost3} by 3 steps in \tabl{noghost}. Note that the trace weighted by $(-)^{\hN}$ is an {\it index}.

The index is very similar to a partition function or a {\it character} of a Virasoro algebra, but there is an important difference. The index sums over the on-shell states only. In flat spacetime, the mass-shell condition can be always satisfied by suitably choosing $k^{\mu}$, so the index takes the same form as the character with weight $(-)^{\hN}$ (apart from a zero-mode contribution $q^{\ap k^2}$). In general, this is not the case though \cite{Asano:2003qb}. 

\begin{table}
\begin{center}
\[
\begin{array}{ccc}
\mbox{tr}_{\obs} \, q^{\Lint} C 
	&\stackrel{\mbox{\footnotesize No-Ghost Theorem}}{=}& 
	\mbox{tr}_{\obs} \, q^{\Lint} \\
&& \\
	\mbox{\footnotesize Step~2.2} \updownarrow 
	&& \updownarrow \mbox{\footnotesize Step~2.1} \\
&& \\
\mbox{tr} \, q^{\Lint} C 
	&\stackrel{\mbox{\footnotesize Step~2.3}}{\leftrightarrow}& 
	\mbox{tr} \, q^{\Lint} (-)^{\hN}
\end{array}
\]
\caption{Strategy to prove the no-ghost theorem.}
\label{table:noghost}
\end{center}
\end{table}
 

\subsection{Step~2.1}

\begin{proof}[Proof of Step~2.1]
At each mass level, states $\varphi_{m}$ in $V_{i}$ are classified into two kinds of representations: BRST singlets $\phi_{\ta} \in V^{\obs}_{i}$ and BRST doublets $(\chi_{a}, \psi_{a})$, where $\chi_{a} = \hQ \psi_{a}$. The ghost number of $\chi_{a}$ is the ghost number of $\psi_{a}$ plus 1. Therefore, $(-)^{\hN}$ causes these pairs of states to cancel in the index and only the singlets contribute:
\begin{eqnarray}
\tr q^{\Lint} (-)^{\hN}
        &=& \mbox{tr}_{\obs} \, q^{\Lint} (-)^{\hN} \\
        &=& \mbox{tr}_{\obs} \, q^{\Lint}.
\end{eqnarray}
We have used the vanishing theorem on the last line.
\end{proof}

\subsection{Step~2.2}

\begin{proof}[Proof of Step~2.2]
At a given mass level, the matrix of inner products among $\ket{\varphi_{m}}$ takes the form
\begin{equation}
\norm{\varphi_{m}}{\varphi_{n}}
=
\left(\begin{array}{c}
        \bra{\chi_{a}} \\
        \bra{\psi_{a}} \\
        \bra{\phi_{\ta}}
\end{array}\right)
\left( \ket{\chi_{b}}, \ket{\psi_{b}}, \ket{\phi_{\tb}} \right)
=
\left( \begin{array}{ccc}
        0 & M & 0 \\
        M^{\dagger} & A & B \\
        0 & B^{\dagger} & D
\end{array} \right).
\label{eq:matrixab}
\end{equation}
We have used $\hat{Q}^\dagger = \hat{Q}$,
$ \norm{\chi}{\chi} = \bra{\chi} \hQ \ket{\psi} = 0 $ and
$ \norm{\chi}{\phi} = \bra{\psi} \hQ \ket{\phi} = 0 $.
If $M$ were degenerate, there would be a state $\chi_{a}$ which is orthogonal to all states in $V_{i}$. Thus, the matrix $M$ should be nondegenerate. (Similarly, the matrix $D$ should be nondegenerate as well.) 
So, a change of basis
\begin{eqnarray}
\ket{\chi'_{a}} &=& \ket{\chi_{a}},
	\nonumber \\
\ket{\psi'_{a}} &=& \ket{\psi_{a}}
	- \frac{1}{2} (M^{-\dagger} A)_{ba} \ket{\chi_{b}},
	\nonumber \\
\ket{\phi'_{\ta}} &=& \ket{\phi_{\ta}}
	- (M^{-\dagger} B)_{b \ta} \ket{\chi_{b}},
\label{eq:transf1}
\end{eqnarray}
sets $A=B=0$. Finally, going to a basis,
\begin{eqnarray}
\ket{\chi''_{a}} &=&
	\frac{1}{\sqrt{2}} (\ket{\chi'_{a}}
	+ M^{-1}_{ba} \ket{\psi'_{b}}), \nonumber \\
\ket{\psi''_{a}} &=&
	\frac{1}{\sqrt{2}} (\ket{\chi'_{a}}
	- M^{-1}_{ba} \ket{\psi'_{b}}), \nonumber \\
\ket{\phi''_{\ta}} &=& \ket{\phi'_{\ta}},
\label{eq:transf2}
\end{eqnarray}
the inner product $\norm{\varphi''_{m}}{\varphi''_{n}}$ becomes block-diagonal:
\begin{equation}
\norm{\varphi''_{m}}{\varphi''_{n}}
=
        \left( \begin{array}{cccccc}
                1 & 0  & 0 \\
                0 & -1 & 0 \\
                0 & 0  & D
        \end{array} \right).
\end{equation}
Therefore, BRST doublets again make no net contribution:
\begin{equation}
\tr q^{\Lint} C
        = \mbox{tr}_{\obs} \, q^{\Lint} C.
\end{equation}
This proves Step~2.2.
\end{proof}

One can indeed check that $M$ and $D$ are nondegenerate. The inner product in $V_i$ is written as the product of inner products in ${\cal F}(\alpha^{0}_{-m}, \psi^{0}_{-r};s, k^0)$, superconformal ghost sector and ${\cal H}_K$. The inner product in ${\cal F}(\alpha^{0}_{-m}, \psi^{0}_{-r};s, k^0)$ is easily seen to be diagonal and nondegenerate.
For the ghost sector, the inner product becomes diagonal and nondegenerate as well by taking the basis 
\renewcommand{\arraystretch}{1.5} 
\be
\begin{array}{rclrcl}
p_{m} &=& \frac{1}{\sqrt{2}}(b_{m}+c_{m}), & 
m_{m} &=& \frac{1}{\sqrt{2}}(b_{m}-c_{m}), \\
p_{r} &=& \frac{1}{\sqrt{2}}(\gam{r}+\bet{r}), & 
m_{r} &=& \frac{1}{\sqrt{2}}(\gam{r}-\bet{r}), 
\end{array}
\ee
whose (anti-)commutation relations are
\be
\begin{array}{rclrclrcl}
\{ p_m, p_n \} &=& \delta_{m+n}, &
\{ p_m, m_n \} &=& 0, &
\{ m_m, m_n \} &=& -\delta_{m+n},
\\
\protect
[ p_r^\dag, p_s ] &=& \delta_{r-s}, &
[ p_r^\dag, m_s ] &=& 0, &
[ m_r^\dag, m_s ] &=& -\delta_{r-s},
\end{array}
\ee
\label{eq:diagonal_basis}
\renewcommand{\arraystretch}{1}%
\noindent
where $\delta_{m+n} = \delta_{m+n, 0}$. For the SCFT $K$ sector, ${\cal H}_K$ is assumed to have a positive-definite inner product. Therefore, the matrix $\norm{\varphi_{m}}{\varphi_{n}}$ is nondegenerate. Consequently, the matrices $M$ and $D$ are also nondegenerate. 

The inner product is nonvanishing only between the states with opposite ghost numbers. Since $D$ is nondegenerate, BRST singlets of opposite ghost number must pair up. We have therefore established the Poincar\'{e} duality theorem as well:
\begin{lemma}[Poincar\'{e} Duality]
BRST singlets of opposite ghost number must pair up, {\it i.e.}, 
\be
H^{\hN}(\hat{\cal H}, \hQ) = H^{-\hN}(\hat{\cal H}, \hQ).
\ee
\label{lemma:poincare}
\end{lemma}
\subsection{Step~2.3}

\begin{proof}[Proof of Step~2.3]
We prove Eq.~\eq{step3} by explicitly calculating the both sides.

In order to calculate the left-hand side of Eq.~\eq{step3}, take an orthonormal basis of definite 
$N^{\rm p}_{m,r}$, $N^{\rm m}_{m,r}$ [the basis \eq{diagonal_basis}], $N^{0}_{m,r}$ and a basis of ${\cal H}_{K} $. Then, 
$C=(-)^{N^{\rm m}_{m} + N^{\rm m}_{r} + N^{0}_{m}+ N^{0}_{r}}$. 
Similarly, for the right-hand side, take an orthonormal basis of definite 
$N^{b}_{m}$, $N^{c}_{m}$, $N^{\beta}_{r}$, $N^{\gamma}_{r}$, $N^{0}_{m,r}$ and an orthonormal basis of ${\cal H}_{K}$. 

%
%

{}From these relations, the left-hand side of Eq.~\eq{step3} becomes
\begin{eqnarray}
\lefteqn{\tr q^{\Lint} C} \nonumber \\
	&=& \left\{ 
		\begin{array}{ll}
			1 &  (\mbox{NS}) \\
			\frac{1}{2} \times 2 & (\mbox{R})
		\end{array}\right\} \times
		q^{\frac{2\nu-9}{16}} \prod_{m,r>0}
			\left( \sum_{N^{\rm p}_{m}=0}^{1}
				q^{m N^{\rm p}_{m}} \right)
			\left( \sum_{N^{\rm m}_{m}=0}^{1}
				q^{m N^{\rm m}_{m}} (-)^{N^{\rm m}_{m}} \right)
			\left( \sum_{N^{0}_{m}=0}^{\infty}
				q^{m N^{0}_{m}} (-)^{N^{0}_{m}} \right)
					\nonumber \\
	&& \mbox{} \hspace{.25in} \times
			\left( \sum_{N^{\rm p}_{r}=0}^{\infty}
				q^{r N^{\rm p}_{r}} \right)
			\left( \sum_{N^{\rm m}_{r}=0}^{\infty}
				q^{r N^{\rm m}_{r}} (-)^{N^{\rm m}_{r}} \right)
			\left( \sum_{N^{0}_{r}=0}^{1}
				q^{r N^{0}_{r}} (-)^{N^{0}_{r}} \right)
			\, \mbox{tr}_{{\cal H}_{K}} \, q^{\LK{0}}
				\nonumber \\
	&=& q^{\frac{2\nu-9}{16}} \prod_{m,r}
			(1+q^{m})(1-q^{m})(1+q^{m})^{-1}
			(1-q^{r})^{-1}(1+q^{r})^{-1}(1-q^{r})
			\, \mbox{tr}_{{\cal H}_{K}} \, q^{\LK{0}}
				\nonumber \\
	&=& q^{\frac{2\nu-9}{16}} \prod_{m,r} \frac{1-q^{m}}{1+q^{r}}
			\, \mbox{tr}_{{\cal H}_{K}} \, q^{\LK{0}}.
\end{eqnarray}
When $q=e^{2\pi i \tau}$, one can rewrite it as
\be
\tr q^{\Lint} C = \left\{ 
		\begin{array}{ll}
		\sqrt{\frac{\eta^3}{\vartheta_{00}(0,\tau)}}\, 
			q^{\LK{0}-\frac{\cK}{16}} & (\mbox{NS}) \\
		\sqrt{\frac{2\eta^3}{\vartheta_{10}(0,\tau)}}\, 
			q^{\LK{0}-\frac{\cK}{16}} & (\mbox{R})
		\end{array}\right.,
\ee
where $\vartheta_{ab} (\nu,\tau)$ is the theta function with characteristics and $\eta(\tau)$ is the Dedekind eta function. The right-hand side becomes
\begin{eqnarray}
\lefteqn{\tr q^{\Lint} (-)^{\hN}} \nonumber \\
	&=& q^{\frac{2\nu-9}{16}} \prod_{m,r>0}
			\left( \sum_{N^{b}_{m}=0}^{1}
				q^{m N^{b}_{m}} (-)^{N^{b}_{m}} \right)
			\left( \sum_{N^{c}_{m}=0}^{1}
				q^{m N^{c}_{m}} (-)^{N^{c}_{m}} \right)
			\left( \sum_{N^{0}_{m}=0}^{\infty}
				q^{m N^{0}_{m}} \right)
					\nonumber \\
	&& \mbox{} \hspace{.25in} \times
			\left( \sum_{N^{\beta}_{r}=0}^{\infty}
				q^{r N^{\beta}_{r}} (-)^{N^{\gamma}_{r}} \right)
			\left( \sum_{N^{\gamma}_{r}=0}^{\infty}
				q^{r N^{\gamma}_{r}} (-)^{N^{\gamma}_{r}} \right)
			\left( \sum_{N^{0}_{r}=0}^{1}
				q^{r N^{0}_{r}} \right)
			\, \mbox{tr}_{{\cal H}_{K}} \, q^{\LK{0}}
				\nonumber \\
	&=& q^{\frac{2\nu-9}{16}} \prod_{m,r}
			(1-q^{m})(1-q^{m})(1-q^{m})^{-1}
			(1+q^{r})^{-1}(1+q^{r})^{-1}(1+q^{r})
			\, \mbox{tr}_{{\cal H}_{K}} \, q^{\LK{0}}
				\nonumber \\
	&=& q^{\frac{2\nu-9}{16}} \prod_{m,r} \frac{1-q^{m}}{1+q^{r}}
			\, \mbox{tr}_{{\cal H}_{K}} \, q^{\LK{0}}.
\end{eqnarray}
This proves Eq.~\eq{step3}.
\end{proof}


%
%

\section*{Acknowledgments}

We would like to thank Vijay Balasubramanian, Mirjam Cvetic, Yang-Hui He, Mitsuhiro Kato, Asad Naqvi, Burt Ovrut and Joe Polchinski for useful discussions. M.N.\ would like to thank theoretical high energy physics group at Univ.\ of Pennsylvania for the kind hospitality where parts of this work was carried out. The research of M.N.\ was supported in part by the Grant-in-Aid for Scientific Research (13740167) and the Oversea Research Fellowship from the Ministry of Education, Culture, Sports, Science and Technology, Japan..

\appendix

\section{Some Basics}\label{sec:appA}

We follow the notations and conventions of Refs.~\cite{Lian:1989cy,Polchinski:rq} (These references occasionally use different conventions; in this case, the conventions of Ref.~\cite{Polchinski:rq} supersede the ones of Ref.~\cite{Lian:1989cy}.) The basic (anti-)commutation relations are
\be
[ \alpha^{\mu}_{m}, \alpha^{\nu}_{n} ] = m \delta_{m+n} \, \eta^{\mu\nu},
\qquad
\{ \psi^{\mu}_{r}, \psi^{\nu}_{s} \} = \delta_{r+s} \, \eta^{\mu\nu},
\ee
\be
\{ b_m, c_n \} = \delta_{m+n},
\qquad
[\gam{r}, \bet{s}] = \delta_{r+s},
\ee
and $\delta_m = \delta_{m,0}$.

The super-Virasoro algebra is given by
\bea
[L_{m}, L_{n}] &=& (m-n) L_{m+n} + \frac{\hc}{8} (m^3-m) \delta_{m+n}, \\
\{G_{r}, G_{s}\} &=& 2 L_{r+s} + \frac{\hc}{8} (4r^2-1) \delta_{r+s}, \\
\protect
[L_{m}, G_r] &=& \frac{m-2r}{2} G_{m+r},
\eea
where $ \hc = 2c/3 $.

The $d=1$ matter part of the super-Virasoro generators are given by
\bea
\Lm{m} &=& -\frac{1}{2} \sum_{n \in {\bf Z}}
                \nor \alpha^{0}_{m-n} \alpha^{0}_{n} \nor
                 - \frac{1}{4} \sum_{r \in {\bf Z}+\nu} (2r-m)
                \nor \psi^{0}_{m-r} \psi^{0}_{r} \nor 
                + \am \delta_m, \\
\Gm{r} &=& - \sum_{n \in {\bf Z}} \alpha^{0}_{n}\psi^{0}_{r-n},
\eea
where $\am = 0 ({\rm NS}), 1/16 ({\rm R})$ and $ \nu = 1/2 ({\rm NS}), 0 ({\rm R}) $. The Noether current for spacetime translation gives $ \alpha^{0}_{0} = \sqrt{2\alpha'} k^{0} $. The superconformal ghost part is given by
\bea
\Lg{m} &=& \sum_{n \in {\bf Z}} (m+n) \nor b_{m-n} c_{n} \nor 
				+ \sum_{r \in {\bf Z}+\nu} \frac{1}{2} (m+2r) 
				\nor \bet{m-r} \gam{r} \nor 
				+ \ag \delta_{m}, \\
\Gg{r} &=& - \sum_{n \in {\bf Z}} 
\{ \frac{1}{2} (2r+n) \bet{r-n}c_n + 2 b_n \gam{r-n} \},
\eea
where $\ag = -1/2 ({\rm NS}), -5/8 ({\rm R})$.

The ghost number operator $\hN$ counts the number of $c$, $\gamma$ minus the
number of $b$, $\beta$ excitations:
\bea
\hN	&=& \sum_{m>0} (c_{-m}b_{m} - b_{-m}c_{m}) 
		- \sum_{r>0} (\gam{-r}\bet{r} + \bet{-r}\gam{r}) 
\\
&=& \sum_{m>0} (N_{m}^{c} - N_{m}^{b})
        + \sum_{r>0} (N_{r}^{\gamma} - N_{r}^{\beta}).
\eea
The operator $\hN$ is related to the standard ghost number operator $ N^{\rm g} $ as 
\be
N^{\rm g}=\hN + c_0 b_0 - (1-2\nu) \gam{-\nu}\bet{\nu} - \nu.
\ee
The ghost zero modes will not matter to our discussion. Note that the operator $\hN$ is also normalized so that $\hN \ket{0_{\rm g}}=0$. ($\ket{0_{\rm g}}$ denotes a ghost ground state. See \app{appB}.)

The BRST operator is
\bea
Q = \sum_{m} c_{-m} (\Lm{m}+\LK{m}) 
+ \sum_r \gam{-r} (\Gm{r}+\GK{r})
        - \sum_{m,n}
        \frac{1}{2} (n-m) \nor b_{-m-n} c_{m} c_{n} \nor 
\nonumber \\
        + \sum_{m,r} \{ \frac{1}{2} (2r-m) \nor \bet{-m-r} c_m \gam{r} \nor
        - \nor b_{-m} \gam{m-r} \gam{r} \nor
        \}
        + \ag c_{0}
\eea
with the part from the unitary SCFT $K$. The BRST operator can be decomposed in terms of ghost zero modes as follows:
\be
Q = 
\left\{
\begin{array}{ll}
\hQ + c_{0} L_{0} + b_{0} M & (\mbox{NS}) \\
\hQ + c_{0} L_{0} + b_{0} M + \gam{0} G_0 + \bet{0} N -\gam{0}^2 b_0 
							& (\mbox{R})
\end{array}
\right.,
\ee
where $ M= -2 \sum_{m>0} (m c_{-m} c_{m} + \gam{-m}\gam{m}) $, $ N = \frac{3}{2} \sum_{r>0} c_{-r}\gam{r}$, and $ \hQ $ is the
collection of the terms in $ Q $ without $ b_{0} $, $ c_{0} $, $\bet{0}$, and $\gam{0}$.

\section{Hilbert Spaces, Ground States, and Inner Products}\label{sec:appB}

We first describe the Hilbert spaces $\Htotal$,
${\cal H}$, and $\hat{\cal H}(k^2)$ more explicitly. 
In particular, we need an appropriate inner product on $\hat{\cal H}(k^2)$
to establish the no-ghost theorem.

The raising operators are $\alpha^{\mu}_{-m}, \psi^{\mu}_{-r}, b_{-m}, c_{-m}, \bet{-r}$, and $\gam{-r}$ $(m, r>0)$. For zero modes, we define that $p_{\mu}, b_0$, and $\bet{0}$ are grouped with the lowering operators and $x^{\mu}, c_0$, and $\gam{0}$ with the raising operators in $\Htotal$. 
In the ghost sector, the ground state is given by $\ket{\downarrow}$, where
\be
b_0 \ket{\downarrow} = 0, \quad
\ket{\uparrow} = c_0 \ket{\downarrow} .
\ee
In the superconformal ghost sector, the ground state is given by

\noindent{\it (i) NS-sector:}
\bea
&& \bet{r} \ket{\frac{1}{2}} = 0, \quad r>0, \\
&& \gam{r} \ket{\frac{1}{2}} = 0, \quad r>0.
\eea

\noindent{\it (ii) R-sector:}
\bea
&& \bet{r} \ket{1} = 0, \quad r \geq 0, \\
&& \gam{r} \ket{1} = 0, \quad r>0.
\eea
The matter R ground states $\ket{s,k}$ are given by a representation
of the gamma matrix algebra of $\psi^{\mu}_0$.
It has 32-dimensional in $d=10$ and 2-dimensional in $d=1$;
they are labeled by $s$.

The Hilbert space ${\cal H}$ is a subspace of $\Htotal$ and it is defined by the condition $b_0=0$ (and $\beta_0=0$ in R-sector) as in Eqs.~\eq{Hilbert_NS} and \eq{Hilbert_R}. So, it is represented as
\be
{\cal H} = \left\{ 
		\begin{array}{ll}
		{\cal F}(\alpha^{\mu}_{-m}, \psi^{\mu}_{-r}; k) \otimes {\cal F}(b_{-m}, c_{-m}, \bet{-r}, \gam{-r}; \downarrow, \frac{1}{2}) 	\otimes {\cal H}_{K} & (\mbox{NS}) \\
		{\cal F}(\alpha^{\mu}_{-m}, \psi^{\mu}_{-r}; s, k) \otimes {\cal F}(b_{-m}, c_{-m}, \bet{-r}, \gam{-r}; \downarrow, 1) \otimes {\cal H}_{K}	& (\mbox{R})
		\end{array}\right.
\ee
Here,  ${\cal F}(\alpha^{\mu}_{-m}, \psi^{\mu}_{-r}; k)$ is a Fock space spanned by 
all $\alpha^{\mu}_{-m}$ and $\psi^{\mu}_{-r}$ ($m, r > 0$)
on the matter ground state $\ket{k} = e^{i k x} \ket{0}$ (and similarly for the others).
A state in ${\cal H}_{K}$ is constructed by Verma modules of $K$
on a highest weight state $\ket{\hK}$. 

In the NS-sector, 
$\hat{\cal H}(k^2)$ is given by imposing $L_0$-condition on ${\cal H}$ :
\begin{equation}
\hat{\cal H}(k^2) =
\left(
{\cal F}(\alpha^{\mu}_{-m}, \psi^{\mu}_{-r}; s, k) \otimes {\cal F}(b_{-m}, c_{-m}, \bet{-r}, \gam{-r}; 0_{\rm g})
\otimes {\cal H}_{K}
\right)^{L_0}.
\end{equation}
where $\ket{0_{\rm g}}$ denotes the ghost ground state $\ket{\downarrow, \frac{1}{2}}$.
The inner product in the space $\hat{\cal H}(k^2)$ is given by
\be
\bra{0,I;k,0_{\rm g}} \ket{0,I';k,0_{\rm g}} = \delta_{II'}
\ee
with the hermiticity property, 
\be
\begin{array}{rclrclrcl}
(\alpha^{\mu}_{m})^\dagger &=& \alpha^{\mu}_{-m}, &
b_{m}^\dagger &=& b_{-m}, &
c_{m}^\dagger &=& c_{-m}, \\
(\psi^{\mu}_{r})^\dagger &=& \psi^{\mu}_{-r}, &
\bet{r}^\dagger &=& -\bet{-r}, &
\gam{r}^\dagger &=& \gam{-r}.
\end{array}
\label{eq:hermiticity_property}
\ee
Here $I$ labels the states of the unitary SCFT $K$.
We take the basis $I$ to be orthonormal. 
Note that the above hermiticity is consistent with the hermiticity 
of the BRST charge $Q$ on the inner product $\bra{}\ket{}$.
The relation of this inner product $\bra{}\ket{}$ with that $\norm{}{}$
in $\Htotal$ is 
\be
\norm{0,I;k,\uparrow,\frac{1}{2}}{0,I';k',\downarrow,\frac{1}{2}}
	= 2 \pi \delta (k^2-k'{}^2) \bra{0;k,0_{\rm g},I} \ket{0;k',0_{\rm g},I'}.
\ee
We write $\bra{\cdots}\ket{\cdots}$ as $\norm{\cdots}{\cdots}$ in this paper.

On the other hand, in the R-sector, the space $\hat{\cal H}(k^2)$ is given by 
$\hat{\cal H}(k^2) = {\cal H}^{L_0,G_0}$.
The $L_0$-condition is the same as the NS-sector and it just 
gives the space ${\cal H}^{L_0}(k^2)$ by imposing 
the condition $\alpha' k^2 + \Lint = 0$ on ${\cal H}$.
To obtain $\hat{\cal H}(k^2)$, we have to impose the condition $G_0=0$ on ${\cal H}^{L_0}$ further. The dimension of $\hat{\cal H}(k^2)$ is half of the space ${\cal H}^{L_0}(k^2)$.
This is verified as follows:
First, note that $G_0$ defines a complex on ${\cal H}^{L_0}$
since $G_0{}^2=0$ in ${\cal H}^{L_0}$.
If $\ket{\phi}$ is a $G_0$-closed state, $G_0 \ket{\phi} = 0$, then 
by using the relation $\{ G_0, {\psi^0_0}/{\alpha^0_0} \} = 1$, 
\be
\ket{\phi} = \bigg\{ G_0, \frac{\psi^0_0}{\alpha^0_0} \bigg\}\ket{\phi}
= G_0 \left( \frac{\psi^0_0}{\alpha^0_0} \ket{\phi} \right).
\ee
Namely, ${\psi^0_0}/{\alpha^0_0}$ is the homotopy operator for $G_0$ and $\ket{\phi}$ is $G_0$-exact. Thus, ${\cal H}^{L_0}$ has no $G_0$-singlets and only $G_0$-doublets exist. The $G_0$-daughter states contribute to $\hat{\cal H}(k^2)$ whereas the $G_0$-parent states do not. Since the number of daughter states is equal to the number of parent states, the space $\hat{\cal H}(k^2)$ has half the states of ${\cal H}^{L_0}(k^2)$. Note that these daughter states can be written as $\hat{\cal H} = G_0 {\cal H}^{L_0}$.

Now we specify the base and the inner product of $\hat{\cal H}(k^2)$ when $d=1$.
In this case, 2-dimensional fermion zero mode vector is 
represented, {\it e.g.}, by $\ket{\pm,k}$ with $\ket{+,k} = \psi^0_0 \ket{-,k}$.
Define a `world-sheet fermion number operator' $f$ which counts the number of all world-sheet fermions {\it without} $\psi_0^0$ in a state $\ket{\phi} \in {\cal H}^{L_0}(k^2)$. [Hence $(-)^{f}\, \ket{\pm}=\ket{\pm}$.]
The exclusion of the zero mode is the difference from the world-sheet fermion number used in the GSO projection.
In ${\cal H}_K$, where states are represented by Verma module of $K$,
the fermion number is defined by the number of $\GK{-r}$'s $(r \geq 0)$.
Using this operator, we divide the space ${\cal H}^{L_0}(k^2)$ into two spaces ${\cal H}^0$ and ${\cal H}^1$ as 
\be
{\cal H}^a = \{ \ket{\phi; \pm, k} |(-)^f = a \},
\ee
where $a=0$ or $1$.
Note that $\mbox{dim } {\cal H}^0 = \mbox{dim } {\cal H}^1 (=\mbox{dim }\hat{\cal H}).$

As remarked earlier, $\hat{\cal H} = G_0 {\cal H}^{L_0}$. One can show that all the states within $G_0 {\cal H}^0$ are independent. Likewise, the states in $G_0 {\cal H}^1$ are independent, and actually $G_0 {\cal H}^0 = G_0 {\cal H}^1$.
So, one can take either $G_0 {\cal H}^0$ or $G_0 {\cal H}^1$ as a base of $\hat{\cal H}$.
We set the non-degenerate inner product of each of these spaces by the inner product of
${\cal H}^0$ or ${\cal H}^1$ which is defined by 
\be
\bra{0,I;s,k,0_{\rm g}}\ket{0,I';s',k,0_{\rm g}} = \delta_{s s'} \delta_{II'}
\label{eq:inner_p_R}
\ee
with the hermiticity property Eqs.~\eq{hermiticity_property}, where $\ket{0_{\rm g}}$ denotes the ghost ground state $\ket{\downarrow, 1}$. Our construction of the Hilbert space and the inner product is essentially the same as Ref.~\cite{Lian:1989cy}.

We can check that the structure (dimension of ${\cal H}^a$, signature, and index) of these two spaces 
$G_0 {\cal H}^a$ $(a=0,1)$ under the above inner product are exactly the same. 
Thus, we perform the calculation concerning the no-ghost theorem 
in \sect{no-ghost} as follows:
\begin{enumerate}
\item First, consider the space ${\cal H}^{L_0}(k^2)$ with the inner product
Eq.~\eq{inner_p_R}.
\item Then, calculate the dimension and the signature in the space ${\cal H}^{L_0}$. 
\item Finally, divide these results by 2.
This gives the correct results on $\hat{\cal H}(k^2)$.
\end{enumerate}
Note that the assumption $ \alpha^{0}_{0} = \sqrt{2\alpha'} k^{0} \ne 0$ is crucial in the above discussion.

\section{Kac Determinant}\label{sec:appC}

For a ${\cal N}=1$ superconformal algebra, a Verma module ${\cal V}(\hc, h)$ consists of all states of the form
\begin{equation}
\ket{{h,\{\lambda\}}} =
G_{-\gam{1}} G_{-\gam{2}} \ldots G_{-\gam{N}} L_{-\lambda_1} L_{-\lambda_2} \ldots L_{-\lambda_M} \ket{h},
\end{equation}
where $ 0 \leq \gam{1} < \gam{2} < \cdots < \gam{N} $ and $ 0 < \lambda_{1} \leq \lambda_{2} \leq \cdots \leq \lambda_{M} $. Here, each $G_{-r}$ acts at most once since $G_{-r}^2 = L_{-2r}+\hc(4r^2-1)/16$.

Consider the matrix of inner products for the states at level $N$:
\be
{\cal M}_{\{\lambda\}, \{\lambda'\}}^{N}(\hc, h) =
\norm{h,\{\lambda\}}{h,\{\lambda'\}},
\qquad \sum_i \gam{i} + \lambda_i = N.
\label{eq:matrix}
\ee
The Kac determinant is then given by
\begin{subequations}
\bea
\mbox{det} [{\cal M}^{N}(\hc, h)]_{\rm NS} &=& 
K_N \prod_{1 \leq rs \leq 2N}
(h-h_{r,s})^{P_{\rm NS}(N-rs/2)},
\\
\mbox{det} [{\cal M}^{N}(\hc, h)]_{\rm R} &=& 
(h-\frac{\hc}{16})^{P_{\rm R}(N)/2} K_N \prod_{1 < rs \leq 2N}
(h-h_{r,s})^{P_{\rm R}(N-rs/2)}, 
\eea
\end{subequations}
where $K_N$ is a positive constant, $r, s =$ positive integer and $r-s=$  even (NS), odd (R). We normalized $\norm{h}{h}=1$. The multiplicity of the roots $P_{\rm NS, R}(k)$ is given by
\begin{subequations}
\bea
\prod_{n=1}^{\infty} \frac{1+q^{n-1/2}}{1-q^n} 
&=& \sum_{k=0}^{\infty} P_{\rm NS}(k) q^k, \\
\prod_{n=1}^{\infty} \frac{1+q^{n-1}}{1-q^n} 
&=& \sum_{k=0}^{\infty} P_{\rm R}(k) q^k.
\eea
\end{subequations}
The zeros of the Kac determinant are
at
\be
h_{r,s} = \frac{\hc-1+\epsilon}{16} + \frac{1}{4} (r \alpha_+ + s \alpha_-)^2,
\ee
where $\epsilon=0$ (NS), 1 (R) and 
\be
\alpha_{\pm} = \frac{1}{4} (\sqrt{1-\hc} \pm \sqrt{9-\hc}).
\ee
In addition, the determinant vanishes at $h=\hc/16$ in the R-sector.

\newpage

\section{Some Useful Commutators}\label{sec:appD}

In this appendix, we collect some useful commutators:

\begin{align}
&[L_{m}, \alpha^{\nu}_{n}] = -n \alpha^{\nu}_{m+n}, 
&&[L_{m}, \psi^{\nu}_{r}] = -\frac{1}{2}(m+2r) \psi^{\nu}_{m+r}, 
\nonumber \\
&[L_{m}, b_{n}] = (m-n) b_{m+n}, 
&&[L_{m}, c_{n}] = (-2m-n) c_{m+n}, 
\nonumber \\
&[L_{m}, \bet{r}] = \frac{1}{2}(m-2r) \bet{m+r}, 
&&[L_{m}, \gam{r}] = - \frac{1}{2}(3m+2r) \gam{m+r}, 
\nonumber \\
&[Q, L_{m}] = 0, 
&& \{ Q, G_{r} \} = 0, \nonumber \\
&[Q, \alpha^{\nu}_{m}] = - \sum_{n} m c_{n} \alpha^{\nu}_{m-n} - \sum_{r} m \gam{-r}\psi^{\nu}_{m+r}, 
&&\{ Q, \psi^{\nu}_{r} \} = \sum_{s} \gam{-s} \alpha^{\nu}_{r+s} - \sum_{s} \frac{1}{2}(s+2r)c_{-s} \psi_{r+s}, 
\nonumber \\
&\{ Q, b_{m} \} = L_{m}, 
&&\{ Q, c_{m} \} = - \sum_{n} n c_{-n} c_{m+n} - \sum_{s} \gam{-s}\gam{m+s}, 
\nonumber \\
&[Q, \bet{r}] = G_{r},
&&[Q, \gam{r}] = - \sum_{s} \frac{1}{2}(3s-r)c_{r-s} \gam{s}, 
\nonumber \\
&[G_r, \alpha^{\nu}_{m}] = - m \psi^{\nu}_{r+m}, 
&&\{ G_r, \psi^{\nu}_{s} \} = \alpha^{\nu}_{r+s}, 
\nonumber \\
&\{ G_r, b_{m} \} = - \frac{1}{2} (2r-m) \bet{r+m}, 
&&\{ G_r, c_{m} \} = -2 \gam{r+m}, 
\nonumber \\
&[G_r, \bet{s}] = -2b_{r+s}, 
&&[G_r, \gam{s}] = \frac{1}{2}(3r+s) c_{r+s}, 
\nonumber \\
&[N^{\rm g}, b_{m}] = -b_{m}, 
&&[N^{\rm g}, c_{m}] = c_{m}, 
\nonumber \\
&[N^{\rm g}, \bet{r}] = -\bet{r}, 
&&[N^{\rm g}, \gam{r}] = \gam{r}, 
\nonumber \\
&[N^{\rm g}, L_{m}] = 0,
&&[N^{\rm g}, Q] = Q,
\nonumber \\
&[N^{\rm g}, G_{r}] = 0. 
\nonumber
\end{align}

\end{document}